# Effects of a Magnetic Field on a Modulated Phase

Bambi Hu[1,2] and Jian Zhou[1,3]

*1. Department of Physics and Centre for Nonlinear and Complex Systems, Hong Kong Baptist University, Kowloon Tong, Hong Kong.*

*2. Department of Physics, University of Houston, Houston, TX 77204, USA.*

*3. Institute of Applied Physics and Computational Mathematics, P.O.Box 8009, 100088 Beijing, P.R.China*

## Abstract

The effect of a magnetic field on a modulated phase is studied. A modulated phase is found to have two critical fields $H_1$ and $H_2$. For a large enough magnetic field, $H_1$ and $H_2$ can be approximated by a linear law. As a result, the minimum magnetic field needed to destroy a modulated phase is a constant. The minimum magnetic field also greatly depends on the order of a commensurate. A very high order commensurate phase and an incommensurate phase are impossible to survive a magnetic field. The behaviour of a magnetoelastic chain in a magnetic field is can be described by a harmless Devil's staircase. The inverse temperature is found to play a role similar to that of a special magnetic field. The deeper physics underlying these new phenomena is the breaking of the left-right symmetry of a phase diagram. As a result a controllable path to a modulated phase is found.

A modulated phase [1], like a commensurate or an incommensurate phase, is an extensively studied subject in condensed matter physics. It can be created by different mechanisms such as a modulated periodical potential [2] or a nonconvex interaction [3]. How a modulated phase behaves in an external field is an interesting problem. In this paper, we propose to study the effects of a magnetic field on a modulated phase.

The magnetoelastic model studied by Marchand and Caille [4] represents a good example of how a modulated phase can be produced by a magnetoelastic coupling. The model deals with a group of elastically coupled particles in which each particle has a classical spin degree of freedom interacting with its first neighbors through an exchange integral $J$ that is a function of the spacing between particles. The coupling between a spin variable and an elastic variable is taken into account by assuming that the exchange integral $J$ depends linearly on the spacing between particles:

$$J(u_{n+1} - u_n) = J_0 - J_1(u_{n+1} - u_n). \tag{1}$$

By tracing over the spin variables of a temperature-dependent free energy function, they change the problem into solving an effective temperature-dependent free energy function of the elastic variables. The equilibrium lattice structure of the system is the ground state of the effective free energy function. Applying the effective potential method proposed by Griffiths and Chou [5], they found a rich phase diagram of this magnetoelastic system. We will use this model for our studies.

The model we study is described by the following magnetic field dependent Hamiltonian

$$\mathcal{H} = \mathcal{H}_e + \sum_{i=1}^{\infty}[J(u_{i+1} - u_i)s_i s_{i+1} + h s_i] \tag{2}$$

$$\mathcal{H}_e = \sum_{i=1}^{\infty}[\frac{1}{2}k_0 u_n^2 + k_1(u_{i+1} - u_i)^2]$$

Here $h$ is the magnetic field. For simplicity, we restrict ourselves to the Ising case, i.e. $s = \pm 1$. The effective free energy functional $F_e$ of the elastic variables is obtained by the following sum over all configurations $\{s_i\}$,

$$e^{-\beta F_e} = e^{-\beta \mathcal{H}_e} \sum_{\{s_i=\pm 1\}} e^{-\beta \sum_{i=1}^{\infty}[J(u_{i+1}-u_i)s_i s_{i+1}+hs_i]} \tag{3}$$

Because $J(u_{i+1}-u_i)$ is dependent on its self-variable, the sum cannot be performed explicitly (see example in Ref [6]). Moreover, a term beyond the nearest neighbor interaction such as the next nearest neighbor interaction will be produced. However, it is known that the nearest neighbor interaction [1–3] is enough to study a modulated phase. Our study also shows that this is the case. So we will drop all terms related to many-body interactions and only keep the two-body interaction term in the following studies. In this way the sum can be performed. We obtain the effective free energy as

$$F_e = \mathcal{H}_e + \sum_{i=1}^{\infty} \psi(J(u_{i+1}-u_i)) \tag{4}$$

$$\psi(J(u_{i+1}-u_i)) = -\beta^{-1} \ln \{ e^{-\beta J(u_{i+1}-u_i)} [\cosh \beta h + (\sinh^2 \beta h + e^{4\beta J(u_{i+1}-u_i)})^{1/2}]\}$$

It obviously reduces to Marchand and Caille's model [4] when $h = 0$. Applying the effective potential method [5], we can study this model numerically in detail.

To use the phase diagrams [4] obtained by Marchand and Caille as a reference, we must adopt the same definitions of the related parameters and scales. In this way $F_e$ can be rewritten as

$$\tilde{F}_e = \sum_{n=1}^{\infty} \frac{1}{2} x_n^2 + \kappa(x_{n+1}-x_n-\nu)^2 + \tilde{\psi}(x_{n+1}-x_n-\nu) \tag{5}$$

$$\tilde{\psi}(x_{n+1}-x_n-\nu) = -\tau \ln \{ e^{(x_{n+1}-x_n-\nu)/\tau} [\cosh H/\tau + (\sinh H/\tau + e^{-4(x_{n+1}-x_n-\nu)/\tau})^{1/2}]\}$$

Here, $\kappa = k_1/k_0$, $x_n = J_1 u_n/(J_1^2/k_0)$, $\tilde{F}_e$, and $\mu$ are, respectively, $F_e$ and $J_0$ scaled in units of $J_1^2/k_0$. $\tau$ and $H$ are defined by $\tau = \beta^{-1}/(J_1^2/k_0)$ and $H = h/(J_1^2/k_0)$. The parameters of a phase space are defined as $\tau'$=$\tau/4$ , $\xi$=$(1+8\kappa)^{-1}$ and $\nu'$=$\nu/(4\xi)$.

Fig. 1 shows the phase diagrams of $\nu'$ vs. $H$ with $\xi = 0.5$ at temperatures $\tau' = 0.05$, 0.28, 0.5 respectively. The phase diagrams with $\tau' = 0.05$, 0.225, 0.3 and $\xi = 0.3$ are shown respectively in Fig. 2. The highest phase indicated in the figures is period 7. In Fig. 1$(a)$ and 1($b$), the colors from red to violet represent the periods 2, 3, 4, 1. Yellow represents period 5, green period 6 and blue period 7. The change of color intensity represents the change of winding number of the same period. For example, in Fig. 1($a$), the phases from left to right are 0/1, 3/7, 1/5 , 1/4, 1/3, 2/5, 1/6, 1/2, 5/6, 3/5, 2/3, 3/4, 4/5, 4/7 respectively. In Fig. 1($c$), yellow, red, green and magenta represent respectively the periods 1, 2, 3 and 4. The area encircled by a black line is the phase with $\omega \neq \omega_J$. To its left, a phase has $\omega_J = 0$ (ferromagnetic or antiferromagnetic interaction) and, to its right, a phase has $\omega_J = \omega$ (antiferromagnetic or ferromagnetic interaction). So the black line is a magnetic separation line. Therefore, a typical phase space consists of three regions: ferromagnetic, antiferromagnetic and an intermediate section.

In general, the magnetic properties of a spin system in a magnetic field is easy to predict. This fact gives us a chance to test the validity of our model. Figs. 1 and 2 show that an intermediate phase changes into an $\omega = \omega_J$ phase with the increase of a magnetic field and eventually all phases on the right of a vertical line $\nu' = 0$ become the $\omega = \omega_J$ phase. The line $\nu' = 0$ separates the ferromagnetic phase from the antiferromagnetic phases. All of these agree reasonably with a spin system in a magnetic field. It is shown that our model is quite reasonable and the effect of a magnetic field has been taken into account correctly.

The most interesting feature of the phase diagrams in Figs. 1 and 2 is that a modulated phase develops linearly with a magnetic field. A phase has two critical magnetic fields. For a specific $\nu'$ and a phase in a $H$ vs. $\nu'$ phase diagram, the lower critical magnetic field $H_1$ is just a magnetic field that a system enters the phase and the higher critical magnetic field is the magnetic field that the system leaves the phase. For a high enough magnetic field, the critical magnetic fields can be approximated by a linear law as

$$H_1 = H_{10} + g(\xi)\nu' \tag{6}$$
$$H_2 = H_{20} + g(\xi)\nu'$$

The $g$ factor is found to be independent of the parameter $\tau'$ and the specific phase. It is dependent on the parameter $\xi$. For example, all the phases in Fig. 1 have the same value of $g = 4$ because $\xi = 0.5$ here. The $g$ value of Fig. 2 is found to be equal to 2.5 for the same reason. As a result $\Delta H = H_{20} - H_{10}$ is a constant for a phase. That means a phase must be destroyed if a magnetic field applied to the phase is higher than this value. The bigger $\Delta H$ of a phase, the more stable it is in a magnetic field. It is found that the higher the order of a commensurate phase, the smaller a magnetic field is needed to destroy it. So it is impossible for a very high order commensurate phase and an incommensurate phase to survive a magnetic field. Therefore, a magnetoelastic chain in a magnetic field is described by a harmless devil's staircase. This is verified further by investigating the behavior of a superdegenerate phase in a magnetic field.

A magnetic field lifts degeneracy. So when a superdegenerate phase is subject to an applied magnetic field, it must shift from one phase to another phase with an increase in the magnetic field. By changing the magnetic field one should be able to reach any phase. A related devil's staircase should be incomplete. However, in this case, the magnetic field destroys high order commensurate and incommensurate phases. So the related devil's staircase must be harmless. Fig. 3 shows the behaviour of an $SDU_1$ superdegenerate phase [4] at the phase space point $\nu' = 0.3501189$, $\xi = 0.133402$, $\tau' = 0.112964$ when a magnetic field is applied to it. Only finite periods have been observed. They form a harmless Devil's staircase. The width of a step in Fig. 3 is just $\Delta H$, the difference between the two critical fields. It decays rapidly with the increase of the period of a phase. The degenerate process is also very interesting. It enters each phase in a cyclic manner with the increase of a magnetic field. There are several cycles in the process. Each cycle is similar. However, only the last one is reached by continuous steps. All phase transitions in the process are first order. Fig. 4 shows a group of typical total effective potentials. They show how the system can enter one phase from another. The sub-minimum related to a subsequent phase is observed. This is a characteristic of a first order phase transition.

At this point a nature question arises: What is the effect of temperature on a modulated phase? The reason to ask the question is because both temperature and magnetic field destroy a modulated phase. Some information can be inferred directly from Eq. (5). Since all the parameters in Eq. 5 are divided by $\tau$, we can intuitively think of $\frac{1}{\tau}$ as playing a role to that of $H$. This point has been shown in Figs. 1 and 2. That is, for a small value of $\tau'$ a phase quickly become linear with a small magnetic field; for a large value of $\tau'$ a larger magnetic field is needed to make it linear. However, the effects of temperature on a modulated phase are quite different from that of a magnetic field. In the $\tau'$ vs. $\nu'$ phase diagrams (see Figs. 1-4 of ref. [4]) obtained by Marchand and Caille, it is not difficult to find that for a large value of $1/\tau'$ the boundary between phases is a vertical line. In this sense the inverse temperature can be treated as a special magnetic field with $g = 0$. So temperature cannot cause a phase to evolve continuously before it reaches a uniform phase. However, when the $g$ factor of a magnetic field is not equal to zero, it can cause a phase to evolve continuously. A magnetic field can be used as a switch of modulated phases. One can start from a uniform phase and go to any desirable phase by adjusting the magnetic field. The basic physics behind these interesting phenomena is the breaking of left-right symmetry of a phase diagram. Temperature does not break the left-right symmetry, and so its $g$ factor is equal to zero. However, a magnetic field breaks the symmetry and it has a nonzero $g$ factor. This is the reason that we can get any modulated phase from a uniform phase by applying a magnetic field to it.

The relationship between temperature and magnetic field is explored further by studying the phase diagram of magnetic field vs. temperature. Fig. 5 shows two typical phase diagrams of $\tau'$ vs. $H$. Phases up to period 7 are indicated. As before, a color represents a phase and a black line is a magnetic separation line. Here only two kinds of magnetic phase are found. One is a phase with $\omega \neq \omega_J$, the other is a phase with $\omega = \omega_J$. It is obvious that there is no definite rule in the phase diagram. This means that the parameters $\tau'$ and $H$ are very different and do not have any relation.

In summary, the effects of a magnetic field on a modulated phase are studied. The phase diagram shows a linear structure in the presence of a magnetic field. The left-right symmetry of a phase diagram is found to be broken by a magnetic field. As a result, a new kind of magnetic modulation is found.

## ACKNOWLEDGMENTS

This work is supported in part by The Research Grant Council RGC/96-97/10 and the Hong Kong Baptist University Faculty Research Grant FRG/95-96/II-09 and FRG/95-96/II-92.

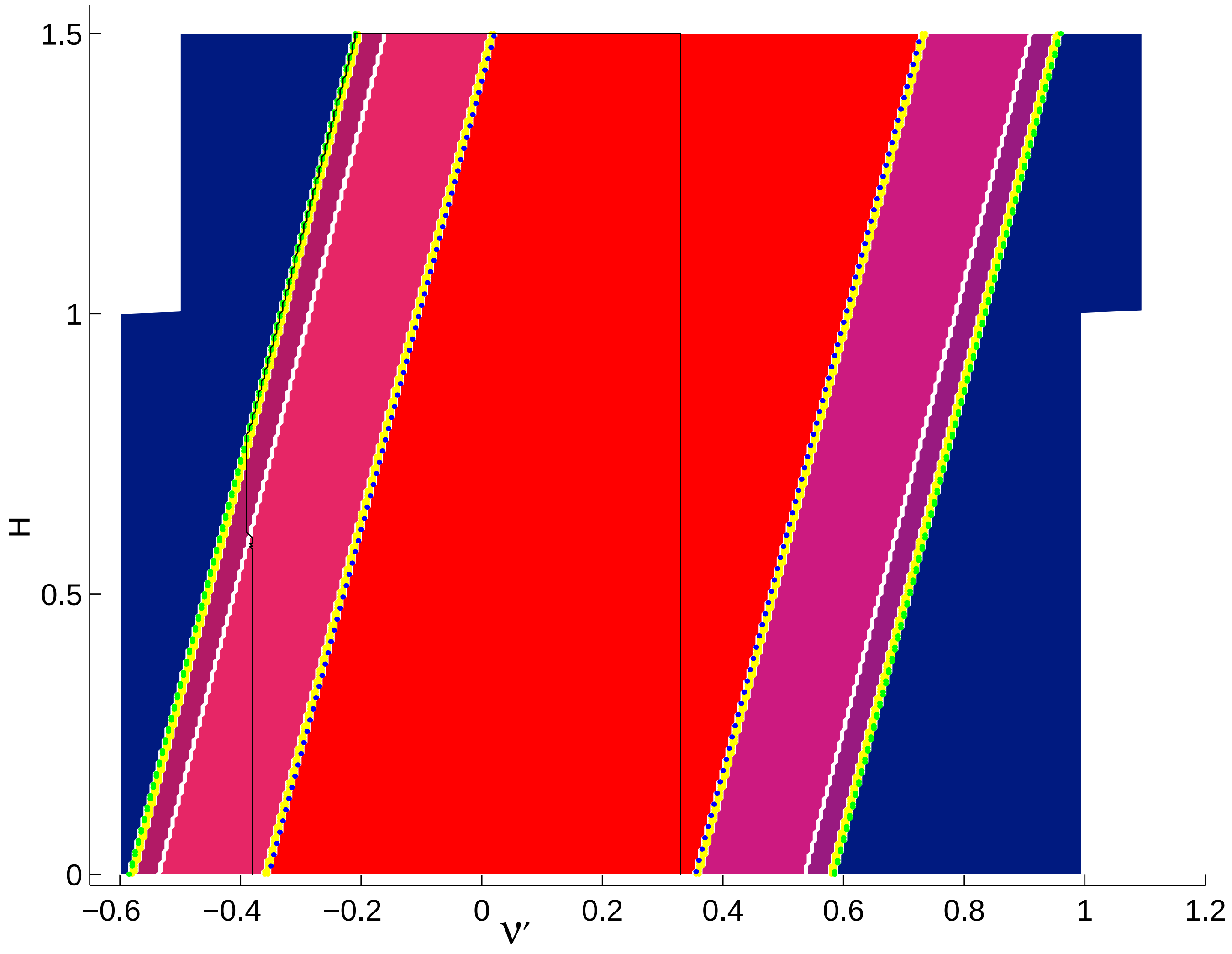

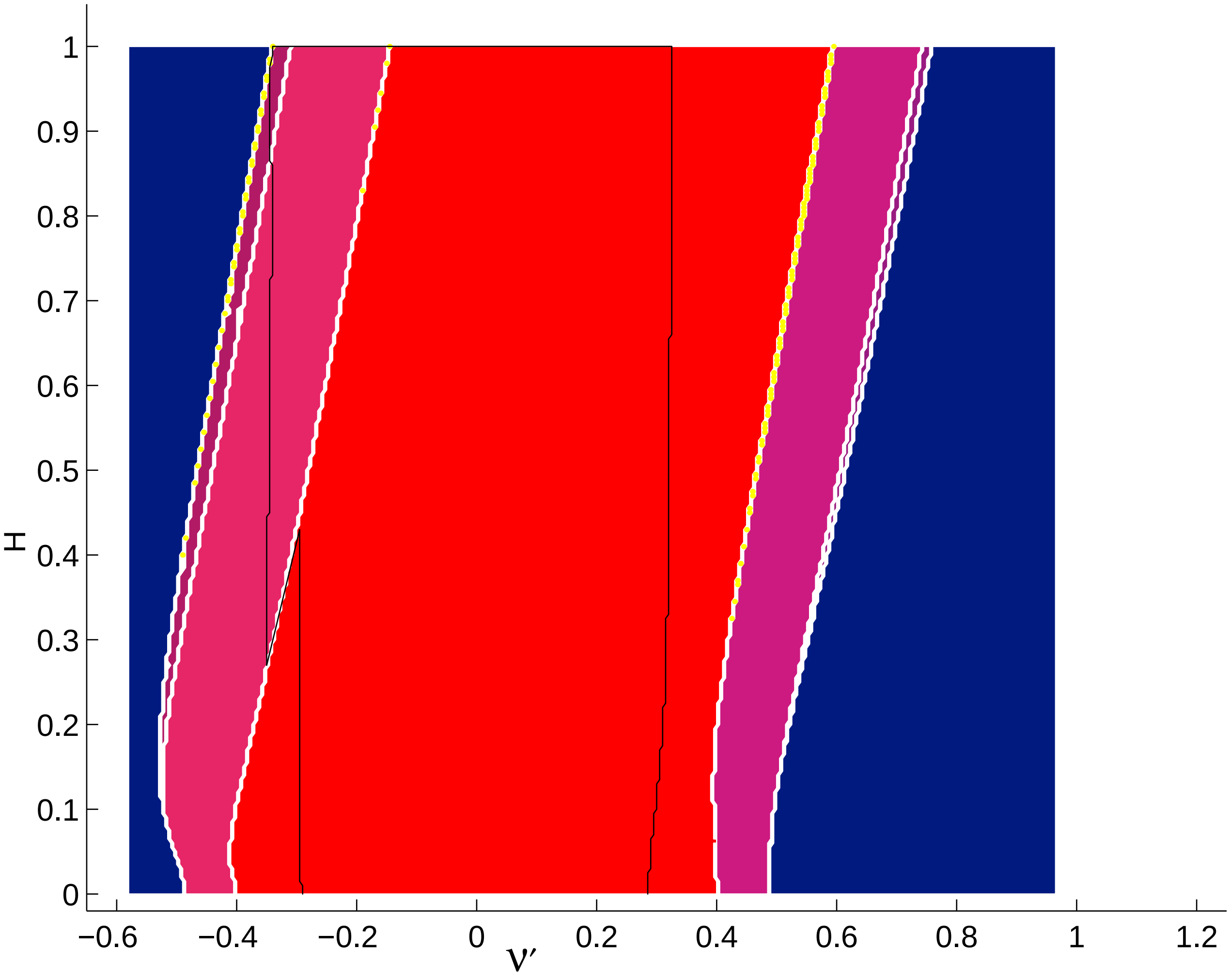

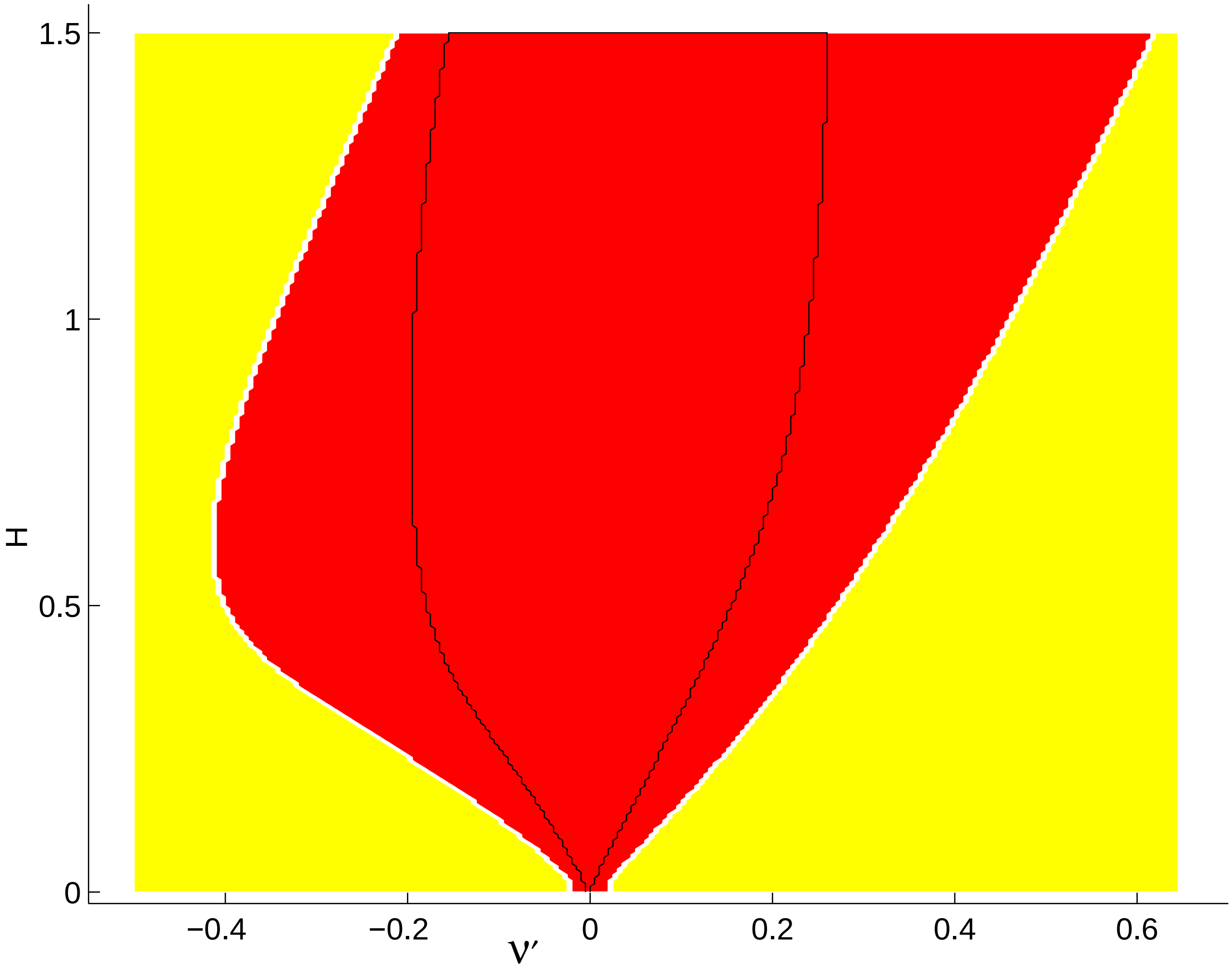

FIG. 1. $H$ vs $\nu'$ phase diagram with $\xi = 0.5$. $(a)$ $\tau'$=0.05, $(b)$ 0.28, $(c)$ 0.5. For $a$ and $b$, the colors from red to violet represent the periods 2, 3, 4, 1. Yellow represents period 5, green period 6 and blue period 7. The change of color intensity represents the change of the winding number of a same period. For example, for $(a)$ the phases from left to right are 0/1, 3/7, 1/5, 1/4, 1/3, 2/5, 1/6, 1/2, 5/6, 3/5, 2/3, 3/4, 4/5, 4/7 respectively. For $(c)$, yellow, red, green and magenta represent respectively the periods 1, 2 ,3 and 4.

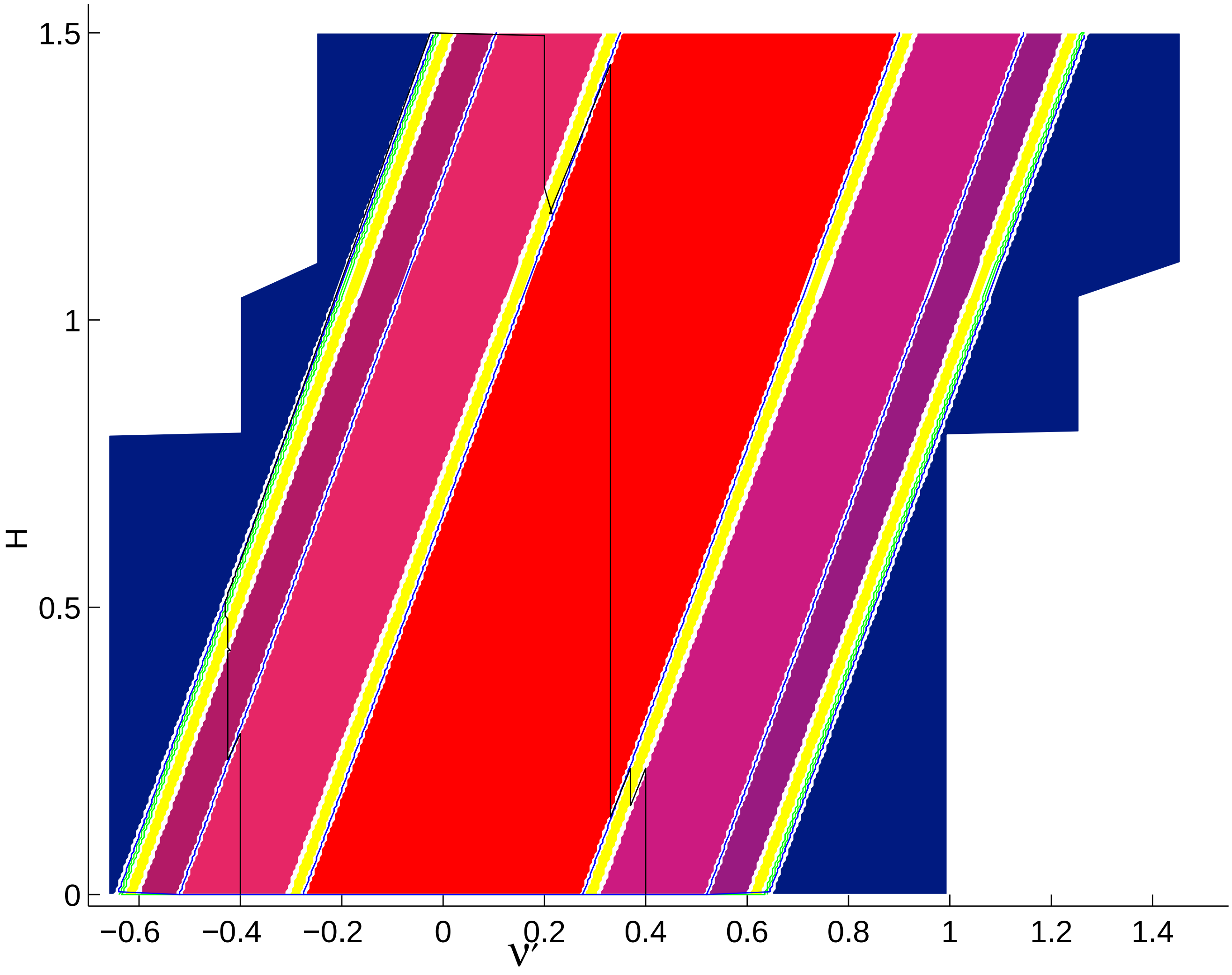

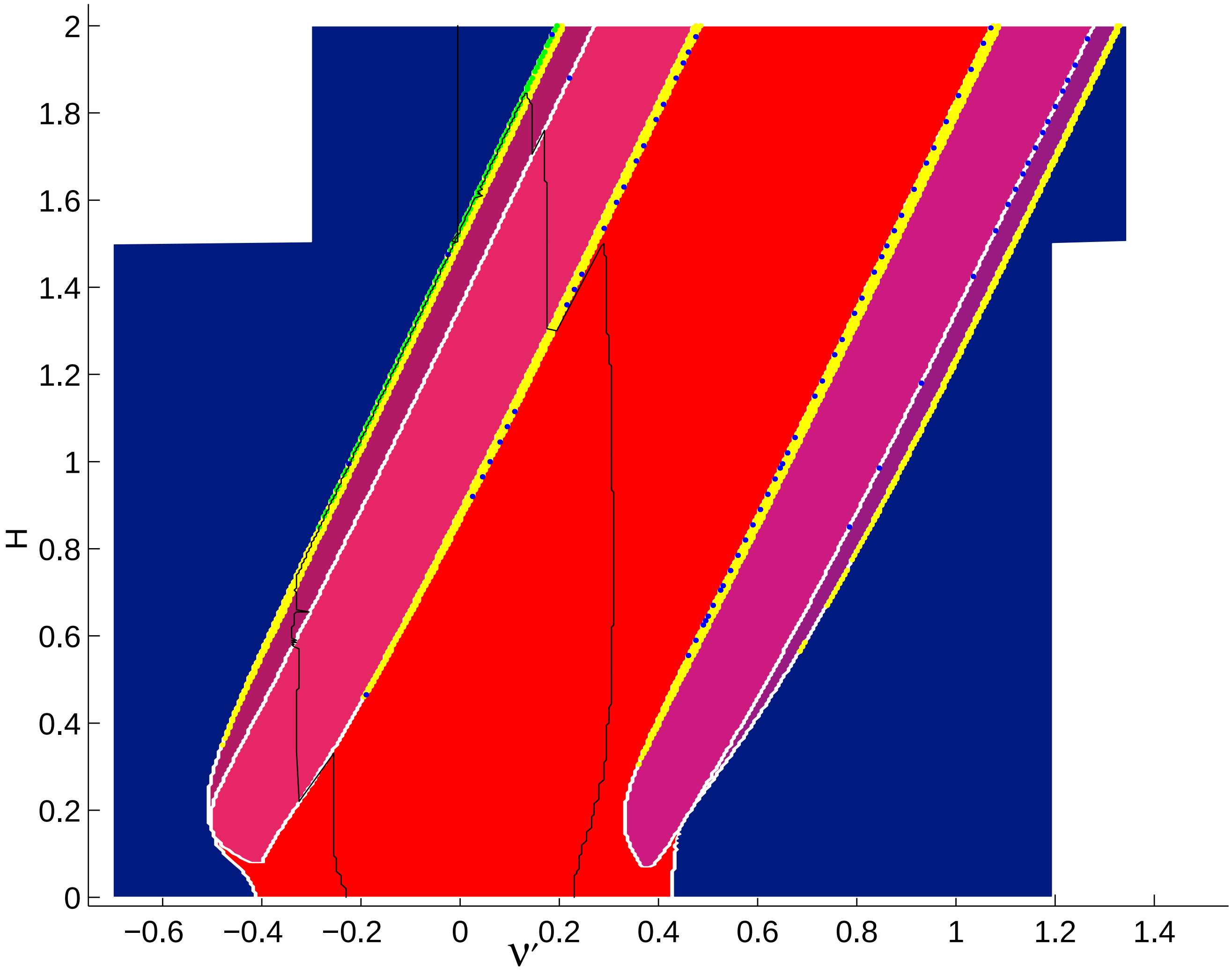

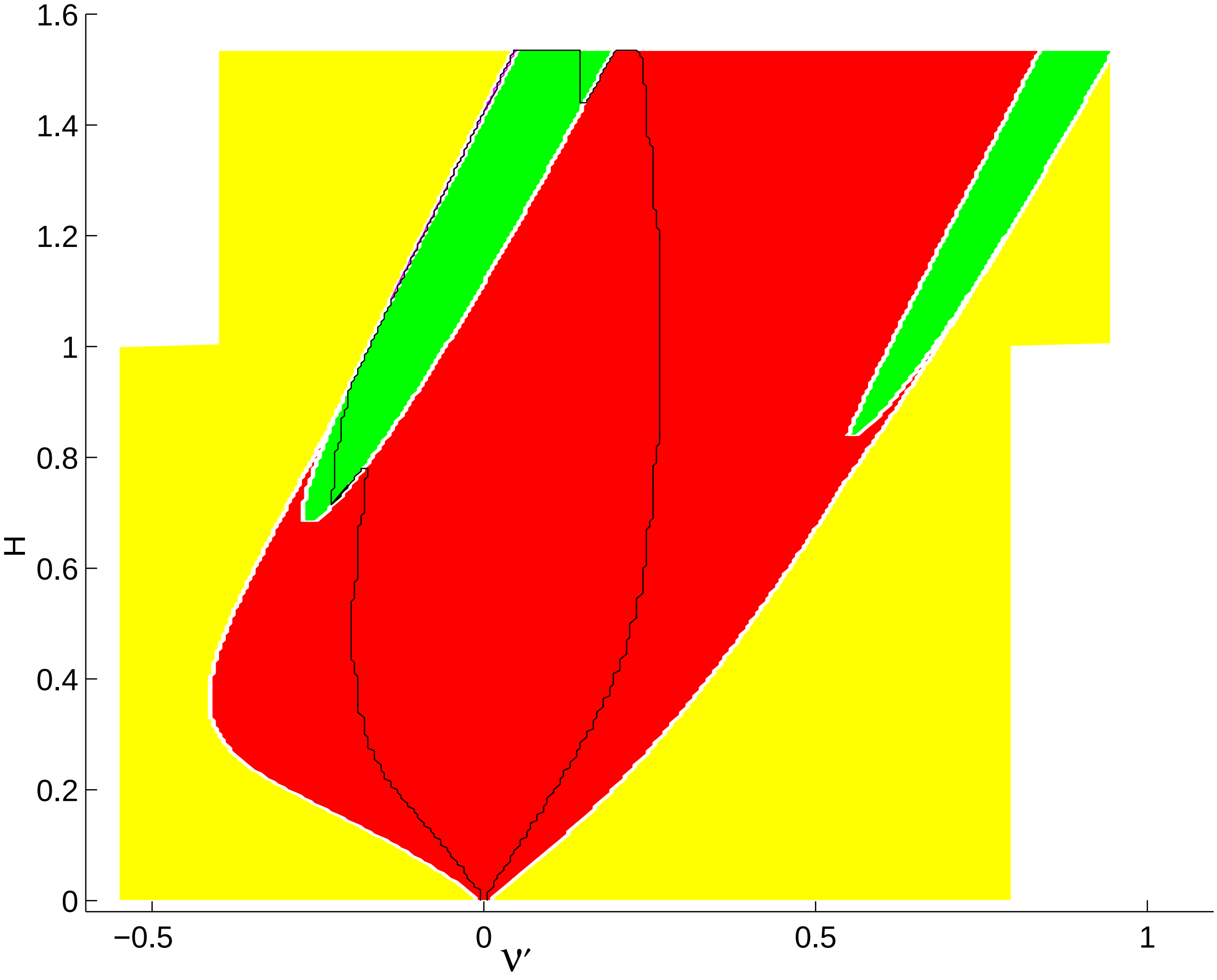

FIG. 2. $H$ vs $\nu'$ phase diagram with $\xi = 0.3$. $(a)$ $\tau'$=0.05, $(b)$ 0.225, $(c)$ 0.3. The color represents the same phase as Fig. 1.

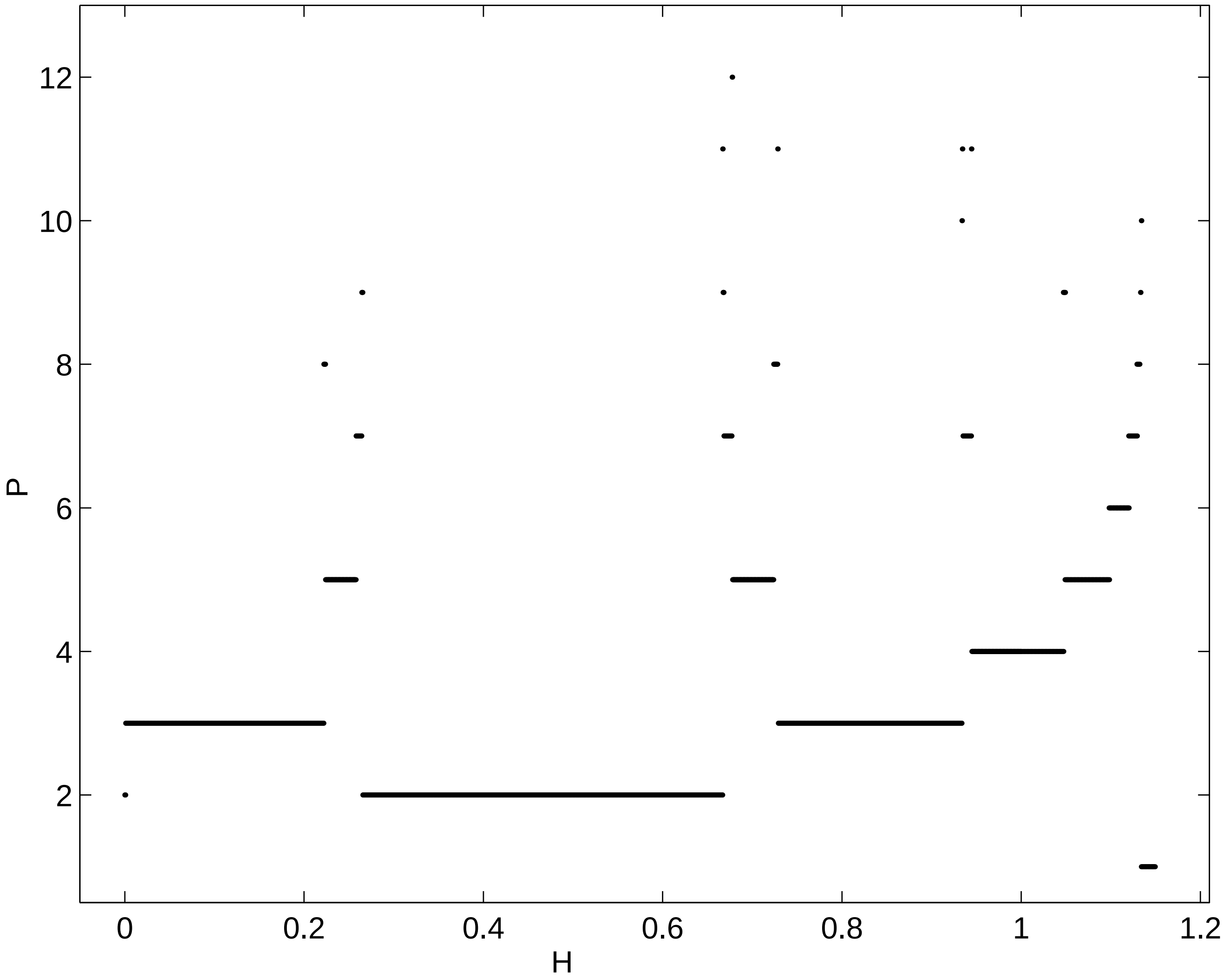

FIG. 3. Devil's staircase of an $SDU_1$ superdegenerate phase.

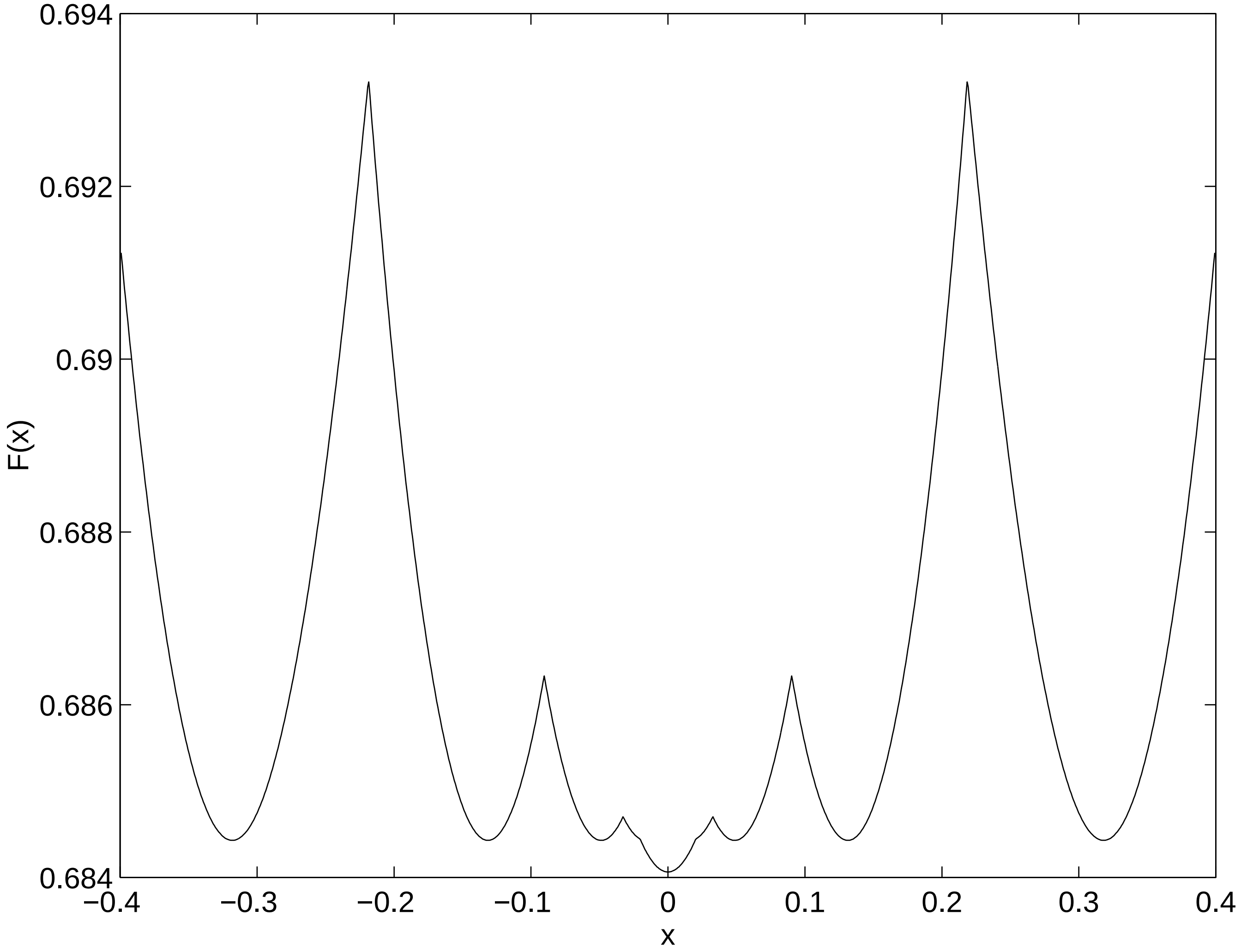

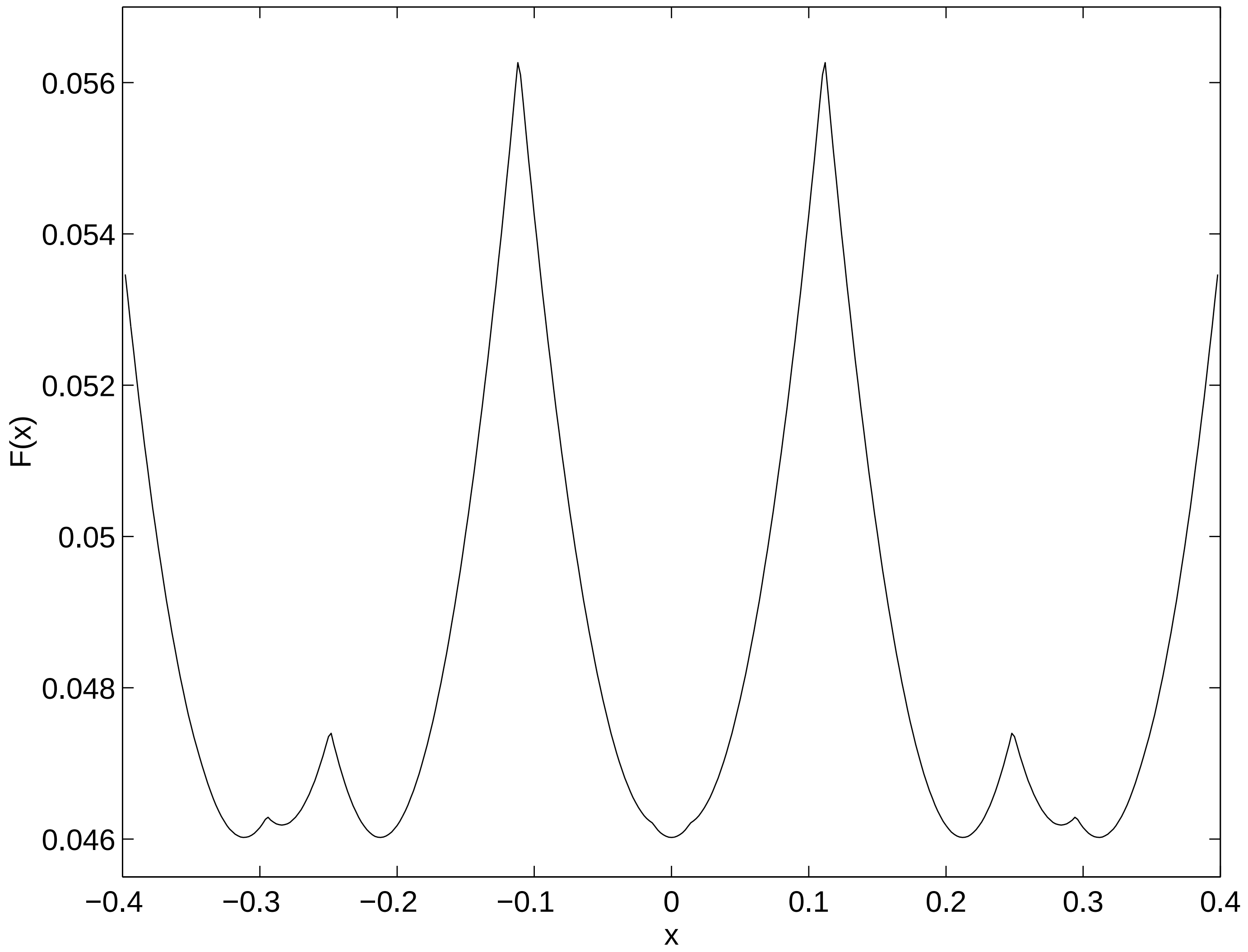

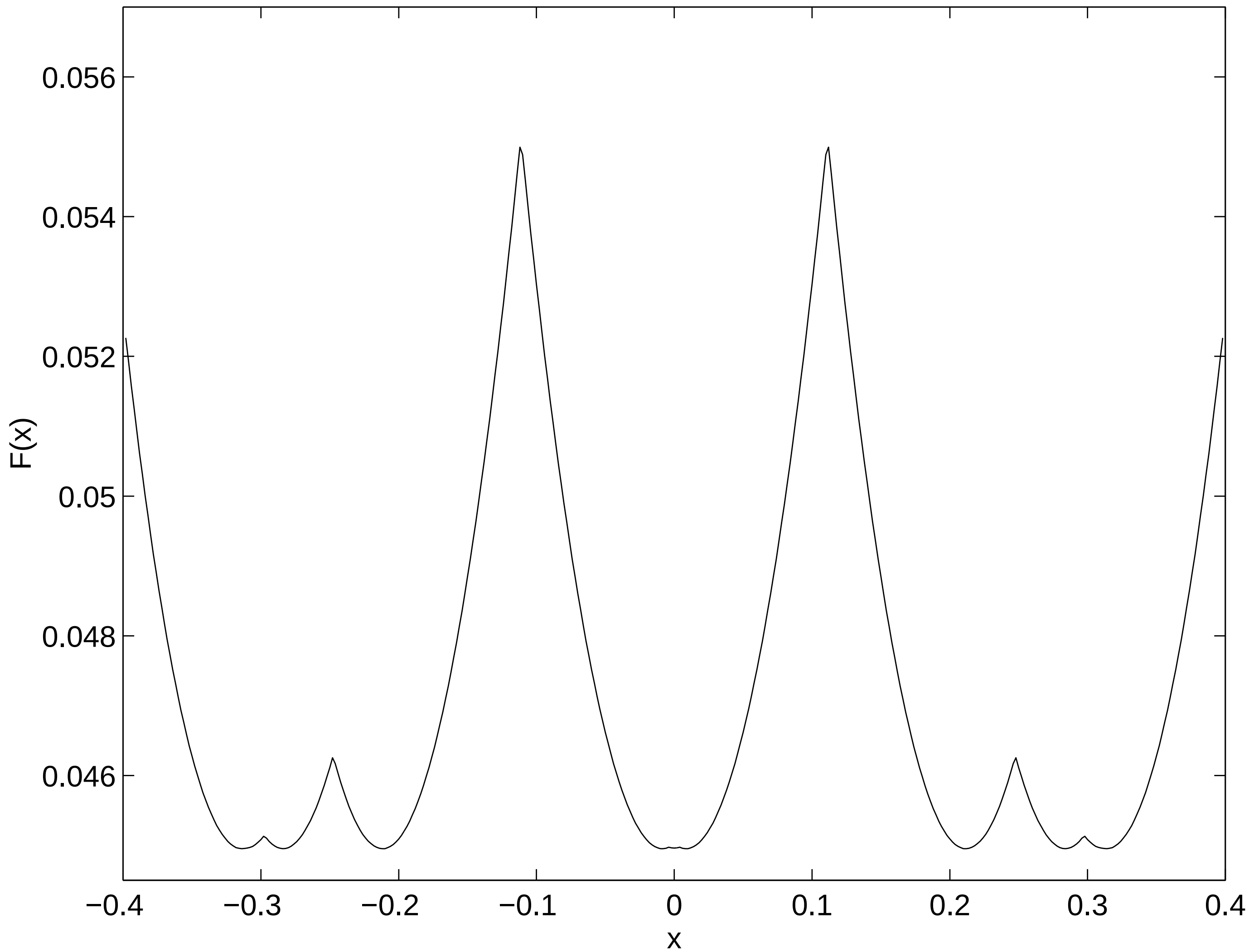

FIG. 4. The change of total effective potential in a phase transition. $(a)$ $H$=1.135, $P$ $(Period)$=1, $(b)$ , $H$=0.225, $P$=5, $(c)$ $H$=0.2235, $P$=8

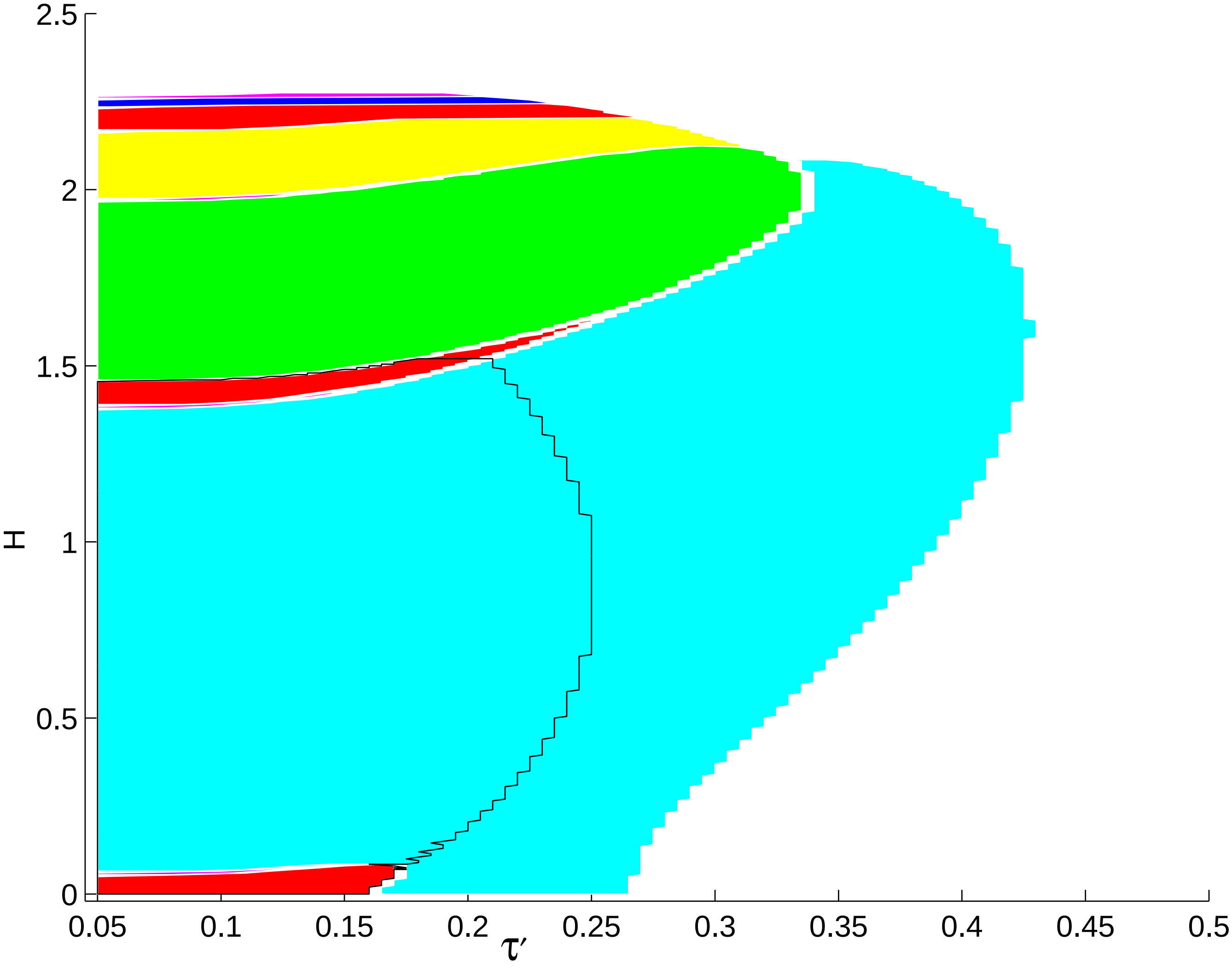

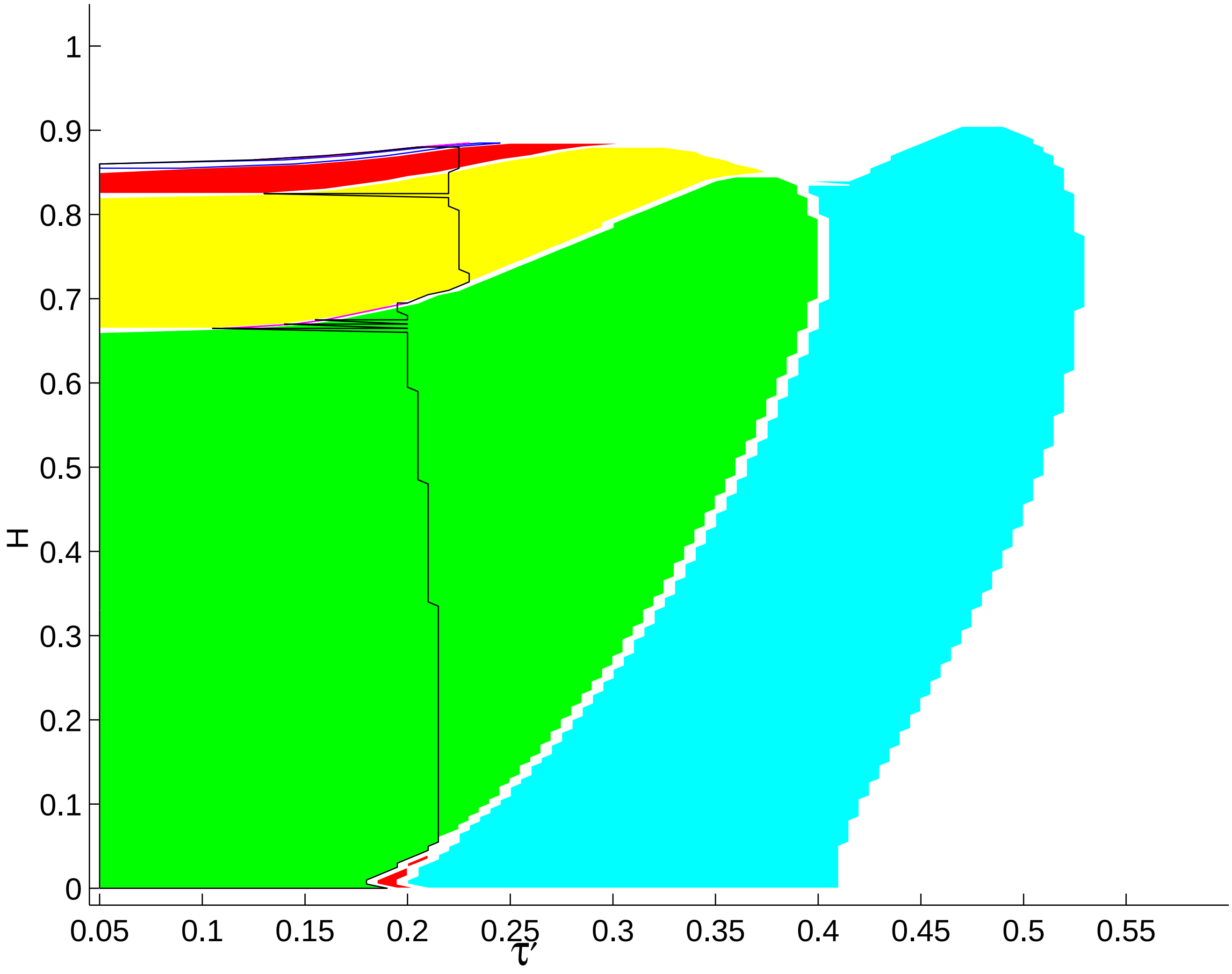

FIG. 5. $H$ vs $\tau'$ phase diagrams with (a) $\xi$=0.3, $\nu'$=0.3 and (b) $\xi$=0.5, $\nu'$=-0.37. The cyan, green, yellow, red, blue, magenta represent respectively the periods 2, 3, 4, 5, 6, 7.